\documentclass{jfm}
\usepackage[T1]{fontenc}
\usepackage[latin9]{inputenc}
\usepackage{soul}
\usepackage{xcolor}
\usepackage{graphicx}
\usepackage{amssymb}

\makeatletter

\providecommand{\tabularnewline}{\\}
\providecolor{lyxadded}{rgb}{0,0,1}
\providecolor{lyxdeleted}{rgb}{1,0,0}

\ifCUPmtlplainloaded \else
  \IfFileExists{amsbsy.sty}
    {\typeout{^^JFound the 'amsbsy' package on the system, using it.^^J}
}

\def\vec#1{\mbox{\boldmath $\mathrm{#1}$}}
\def\tilde#1{\mbox{\boldmath $\mathsf{#1}$}}
\def\RE{\textit{Re}}
\def\Rm{\textit{Rm}}
\def\Ha{\textit{Ha}}

\affiliation{Applied Mathematics Research Centre, Department of Mathematical Sciences, \\       Coventry University, Priory Street, Coventry CV1 5FB, UK}
\pubyear{2009}
\volume{0}

\begin{document}

\title[Linear stability of Hunt's flow]{Linear stability of Hunt's flow}
\author[J. Priede, S. Aleksandrova and S. Molokov]
{J\ls \=A\ls N\ls I\ls S\ns P\ls R\ls I\ls E\ls D\ls E,\ns 
S\ls V\ls E\ls T\ls L\ls A\ls N\ls A\ns A\ls L\ls E\ls K\ls S\ls A\ls N\ls D\ls R\ls O\ls V\ls A\\
and\ns 
S\ls E\ls R\ls G\ls E\ls I\ns M\ls O\ls L\ls O\ls K\ls O\ls V}
\date{}

\maketitle
\begin{abstract}
We analyse numerically the linear stability of the fully developed
flow of a liquid metal in a square duct subject to a transverse magnetic
field. The walls of the duct perpendicular to the magnetic field are
perfectly conducting whereas the parallel ones are insulating. In
a sufficiently strong magnetic field, the flow consists of two jets
at the insulating walls and a near-stagnant core. We use a vector
stream function formulation and Chebyshev collocation method to solve
the eigenvalue problem for small-amplitude perturbations. Due to the
two-fold reflection symmetry of the base flow the disturbances with
four different parity combinations over the duct cross-section decouple
from each other. Magnetic field renders the flow in a square duct
linearly unstable at the Hartmann number $\Ha\approx5.7$ with respect
to a disturbance whose vorticity component along the magnetic field
is even across the field and odd along it. For this mode, the minimum
of the critical Reynolds number $\RE_{c}\approx2018,$ based on the
maximal velocity, is attained at $\Ha\approx10.$ Further increase
of the magnetic field stabilises this mode with $\RE_{c}$ growing
approximately as $\Ha.$ For $\Ha>40,$ the spanwise parity of the
most dangerous disturbance reverses across the magnetic field. At
$\Ha\approx46$ a new pair of most dangerous disturbances appears
with the parity along the magnetic field being opposite to that of
the previous two modes. The critical Reynolds number, which is very
close for both of these modes, attains a minimum, $\RE_{c}\approx1130,$
at $\Ha\approx70$ and increases as $\RE_{c}\approx91\Ha^{1/2}$ for
$\Ha\gg1.$ The asymptotics of the critical wavenumber is $k_{c}\approx0.525\Ha^{1/2}$
while the critical phase velocity approaches $0.475$ of the maximum
jet velocity. 
\end{abstract}

\section{Introduction}

Application of a strong magnetic field to a flow of an electrically
conducting fluid is associated primarily with two effects. First,
the magnetic field acts on the mean flow profile often creating inflexion
points (\cite{Kak64}), shear layers (\cite{Leh52}) and jets (\cite{Hun65}),
thus destabilising the otherwise stable flow. Secondly, strong magnetic
field tends to damp three-dimensional perturbations making them anisotropic,
aligned with the magnetic field, and to transform them into quasi-two-dimensional
structures (\cite{Mof67,Dav95}). There are flows, which combine the
effect of high electromagnetic damping in some flow regions, high
transverse shear in other regions, such as jets, and moderate stretching
along the magnetic field. Instabilities and turbulence may strongly
affect the transfer of momentum, heat, and mass in such liquid metal
flows which are of major importance for various industrial applications
ranging from metallurgy and semiconductor crystal growth (\cite{Dav99})
to the designs of fusion reactors with magnetic confinement (\cite{Buh07}).

\begin{figure}
\begin{centering}
\includegraphics[bb=80bp 55bp 350bp 280bp,clip,width=0.5\textwidth]{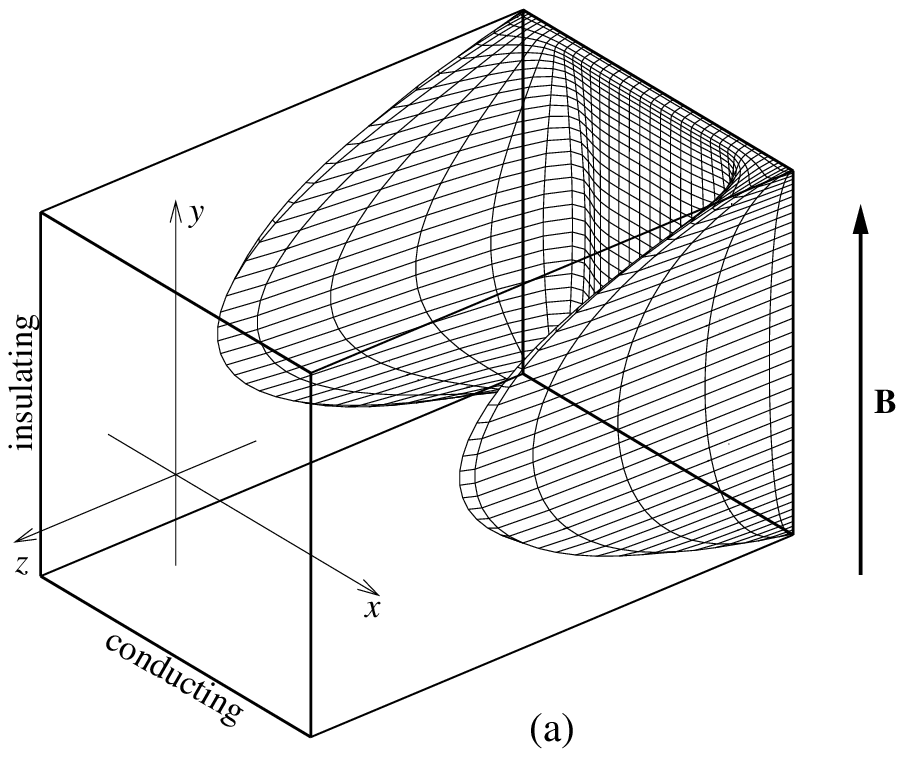}\includegraphics[bb=130bp 90bp 320bp 255bp,clip,width=0.5\textwidth]{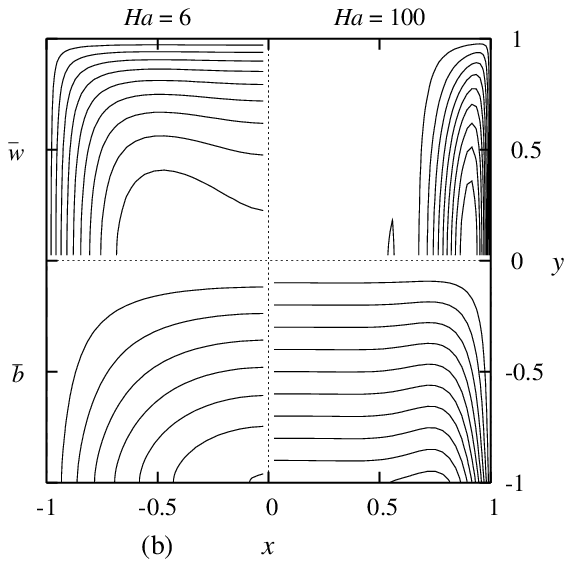}\\
\includegraphics[width=0.5\textwidth]{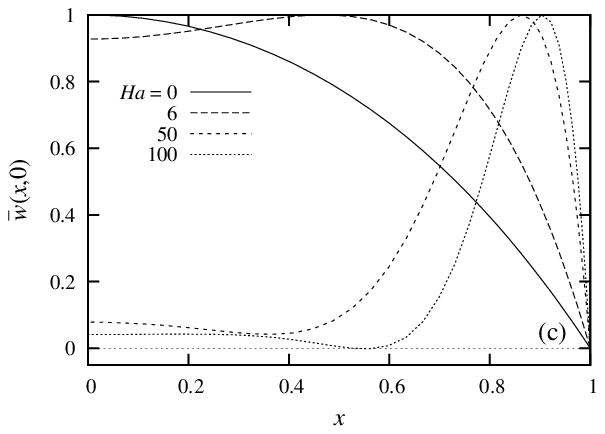} 
\par\end{centering}

\caption{\label{fig:sketch} Sketch to the formulation of the problem with
the base flow profile for $\Ha=100$ (a), isolines of base flow $(y>0)$
and electric current lines $(y<0)$ for $\Ha=6$ $(x<0)$ and $\Ha=100$
$(x>0)$ shown in the respective quadrants of duct cross-section (b)
and the base flow velocity profiles at $y=0$ for $\Ha=0,6,50,100$
(c). }

\end{figure}

Here we will be concerned with linear stability of the fully developed,
isothermal, magnetohydrodynamic (MHD) flow in a constant-area square
duct with a pair of perfectly electrically conducting and another
pair of perfectly insulating walls in the presence of a strong magnetic
field. The field is parallel to the insulating walls and perpendicular
to the conducting ones, and such a flow is known as the Hunt's flow
(\cite{Hun65}). For strong magnetic fields this flow has a pair of
characteristic sidewall jets developing along the insulating walls
while the velocity in the core of the duct is significantly reduced
(see figure \ref{fig:sketch}). These effects are due to the pattern
of the electric currents, shown in figure \ref{fig:sketch}(b) for
$y<0.$ In the core of the duct the electric currents are induced
in the direction transverse to the magnetic field and, thus, the resulting
electromagnetic force nearly balances the pressure gradient driving
the flow. At the insulating side walls, the electric current turns
almost parallel to the magnetic field and, thus, the electromagnetic
braking force in these regions is significantly reduced. As a result,
the applied pressure gradient is balanced there mainly by the viscous
shear, and the flow protrudes through the magnetic field in thin jets
along the sidewalls. In the limit of a very strong magnetic field,
the jets, which are of fundamental importance for liquid metal blankets
in fusion reactors (see e.g. \cite{Sti-etal96,MolBuh94,Mol93}) carry
almost all of the volume flux in the so-called parallel layers. Such
a velocity profile is highly unstable, as has been confirmed experimentally
by \cite{GelDorSch71} and \cite{PlaFre72}. Linear stability analysis
of Hunt's flow has been attempted by \cite{Fuj89} by assuming two-dimensional
mean velocity profile and two-dimensional disturbances, both in the
mid-plane of the duct transverse to the field. His results are of
limited interest owing to the three-dimensional nature of both the
mean profile and the disturbances. A three-dimensional linear stability
analysis of a single sidewall jet has been carried out by \cite{TinWalReePic91}.
They consider the flow in a rectangular duct with thin conducting
walls in the presence of a strong transverse magnetic field. Although
\cite{TinWalReePic91} assume the wall conductance ratio to be small,
the magnetic field is supposed to be so strong that the relative conductance
of both Hartmann and parallel layers is even smaller than that of
the walls. As a result, the induced electric current passes from the
core region directly through the parallel layer into the side wall
and then to close through normal walls back to the core region. Thus,
both the sidewalls and Hartmann walls are treated by \cite{TinWalReePic91}
as effectively well-conducting boundaries. Direct numerical simulation
of this flow has been undertaken by \cite{Muc00} for the Reynolds
number significantly above the linear stability threshold for the
side layers predicted by \cite{TinWalReePic91}.

Here, we present the results of the three-dimensional linear stability
analysis of Hunt's flow in a square duct, which according to \cite{TatYos90}
is linearly stable in the absence of a magnetic field. We show that
the instability is far more complex than predicted by \cite{TinWalReePic91}
and \cite{Fuj89} using asymptotic theory and two-dimensional approximation,
respectively. A magnetic field of moderate strength is found to render
the flow linearly unstable with respect to two pairs of antisymmetric
streak-like perturbations of the axial velocity concentrated in the
middle part of the duct. The most dangerous perturbation is essentially
3D with the component of vorticity along the magnetic field being
even and odd function across and along the magnetic field, respectively.
This instability is associated with the appearance of two velocity
minima in the centre of the duct which at stronger magnetic fields
develop into the sidewall jets. As the magnetic field strength increases,
another essentially 3D instability mode with the opposite parity across
the magnetic field appears. The critical wavelength of both these
modes exceeds the width of the duct several times even in relatively
strong magnetic fields. At the same time, the phase velocity strongly
correlates with the maximum jet velocity. In a sufficiently strong
magnetic field, the critical Reynolds number, based on the maximum
velocity, increases nearly directly with the magnetic field strength.
As the sidewall jets develop, two new, much more unstable modes appear
with the parity along the magnetic field opposite to that of two previous
modes. The critical Reynolds number, which is almost the same for
the last two modes, increases at high magnetic fields inversely with
the side layer thickness while the critical wavelength reduces directly
with the thickness. 

The paper is organised as follows. In Section $2$ below we formulate
the problem. Numerical method is outlined and verified in $\S3$ and
numerical results are discussed in $\S4.$ Section 5 summarises and
concludes the paper.

\section{Problem formulation}

Consider a flow of an incompressible, viscous, electrically conducting
liquid with density $\rho,$ kinematic viscosity $\nu$ and electrical
conductivity $\sigma$ driven by a constant gradient of pressure $p$
applied along a straight duct of rectangular cross-section with half-width
$d$ and half-height $h$ subject to a homogeneous transverse magnetic
field $\vec{B}.$ The walls of the duct perpendicular to the magnetic
field are perfectly conducting whereas the parallel ones are insulating.

The velocity distribution of the flow is governed by the Navier-Stokes
equation\begin{equation}
\partial_{t}\vec{v}+(\vec{v}\cdot\vec{\nabla})\vec{v}=-\frac{1}{\rho}\vec{\nabla}p+\nu\vec{\nabla}^{2}\vec{v}+\frac{1}{\rho}\vec{f},\label{eq:NS}\end{equation}
 where $\vec{f}=\vec{j}\times\vec{B}$ is the electromagnetic body
force involving the induced electric current, which is governed by
the Ohm's law for a moving medium \begin{equation}
\vec{j}=\sigma(\vec{E}+\vec{v}\times\vec{B}).\label{eq:Ohm}\end{equation}
 The flow is assumed to be sufficiently slow so that the induced magnetic
field is negligible with respect to the imposed one, implying the
magnetic Reynolds number $\Rm=\mu_{0}\sigma v_{0}d\ll1,$ where $\mu_{0}$
is the permeability of vacuum and $v_{0}$ is the characteristic velocity
of the flow. In addition, we assume that the characteristic time of
velocity variation is much longer than the magnetic diffusion time
$\tau_{m}=\mu_{0}\sigma d^{2}$ that allows us to use the quasi-stationary
approximation, according to which $\vec{E}=-\vec{\nabla}\phi,$ where
$\phi$ is the electrostatic potential. The velocity and current satisfy
the mass and charge conservation $\vec{\nabla}\cdot\vec{v}=\vec{\nabla}\cdot\vec{j}=0.$
Applying the latter to the Ohm's law (\ref{eq:Ohm}) yields \begin{equation}
\vec{\nabla}^{2}\phi=\vec{B}\cdot\vec{\omega},\label{eq:phi}\end{equation}
 where $\vec{\omega}=\vec{\nabla}\times\vec{v}$ is vorticity. At
the walls of the duct $S$, the normal $(n)$ and tangential $(\tau)$
velocity components satisfy the impermeability and no-slip boundary
conditions, namely $\left.v_{n}\right\vert _{s}=0$ and $\left.v_{\tau}\right\vert _{s}=0.$
The conditions for the electric current at insulating and perfectly
conducting walls are $\left.j_{n}\right\vert _{s}=0$ and $\left.j_{\tau}\right\vert _{s}=0,$
respectively. Boundary conditions for the current and velocity applied
to Ohm's law result in $\left.\partial_{n}\phi\right\vert _{s}=0$
and $\left.\phi\right\vert _{s}=\textrm{const}$ for $\phi$ at insulating
and perfectly conducting walls, respectively.

We employ the Cartesian coordinates with the origin set at the centre
of the duct and with $x$, $y$ and $z$ axes directed along the width,
height and length of the duct, respectively, as shown in figure \ref{fig:sketch},
with the velocity distribution given by $\vec{v}=(u,v,w).$ The problem
admits a purely rectilinear base flow with a single velocity component
along the duct $\bar{\vec{v}}=(0,0,\bar{w}(x,y))$ which is shown
in figure \ref{fig:sketch}(a) for $\Ha=100.$ In the following, all
variables are non-dimensionalised by using the maximum velocity $\bar{w}_{0}$
and the half-width of the duct $d$ as the velocity and length scales,
while the time, pressure, magnetic field and electrostatic potential
are scaled by $d^{2}/\nu,$ $\rho\bar{w}_{0}^{2},$ $B=\left\vert \vec{B}\right\vert $
and $\bar{w}_{0}dB,$ respectively. Note that we use the maximum rather
than average velocity as the characteristic scale because the stability
of this flow is determined by the former as discussed in the following.

Base flow can more conveniently be described using the $z$-component
of the induced magnetic field $\bar{b}$ instead of the electrostatic
potential $\bar{\phi}$ (\cite{Mor90}). This temporal change of variables
does not affect the following linear stability analysis which requires
the base flow profile but not the electrostatic potential. Then the
governing equations for the base flow take the form \begin{eqnarray}
\vec{\nabla}^{2}\bar{w}+\Ha\partial_{y}\bar{b} & = & \bar{P,}\label{eq:wbar}\\
\vec{\nabla}^{2}\bar{b}+\Ha\partial_{y}\bar{w} & = & 0,\label{eq:bbar}\end{eqnarray}
where $\Ha=dB\sqrt{\sigma/(\rho\nu)}$ is the Hartmann number and
$\bar{b}$ is scaled by $\mu_{0}\sqrt{\sigma\rho\nu^{3}}/d$ . Note
that the isolines of $\bar{b}$ represent electric current lines which
are shown in the bottom part of figure \ref{fig:sketch}(b) for $\Ha=6$
and $\Ha=100.$ The dimensionless constant axial pressure gradient
$\bar{P},$ which drives the flow, is determined from the normalisation
condition $\bar{w}_{\max}=1.$ The velocity satisfies the no-slip
boundary condition $\bar{w}=0$ at $x=\pm1$ and $y=\pm A,$ where
$A=h/d$ is the aspect ratio, which is equal to $1$ for the square
cross-section duct considered in this study. The boundary conditions
for the induced magnetic field at insulating and perfectly conducting
walls are $\bar{b}=0$ $(x=\pm1)$ and $\partial_{y}\bar{b}=0$ $(y=\pm A),$
respectively. The base flow is obtained numerically by the Chebyshev
collocation method which is described and validated in the next section.

In order to satisfy the incompressibility constraint $\vec{\nabla}\cdot\vec{v}=0$
for the flow perturbation, we are looking for the velocity distribution
as $\vec{v}=\vec{\nabla}\times\vec{\psi},$ where $\vec{\psi}$ is
a vector stream function. The vector stream function as the magnetic
vector potential $\vec{A}$ in electrodynamics is defined up to a
gradient of an arbitrary function which added to $\vec{\psi}$ does
not change $\vec{v}$. In order to eliminate this ambiguity, we impose
an additional constraint on $\vec{\psi}$\begin{equation}
\vec{\nabla}\cdot\vec{\psi}=0,\label{eq:divpsi}\end{equation}
 which is analogous to the Coulomb gauge for $\vec{A}$ (\cite{Jac98}).
This gauge, similarly to the incompressibility constraint for $\vec{v},$
leaves only two independent components of $\vec{\psi.}$ 

The pressure gradient is eliminated by applying \textit{curl} to (\ref{eq:NS})
which yields two dimensionless equations for $\vec{\psi}$ and $\vec{\omega}$
\begin{eqnarray}
\partial_{t}\vec{\omega} & = & \vec{\nabla}^{2}\vec{\omega}-\RE\vec{g}+\Ha^{2}\vec{h},\label{eq:omeg}\\
0 & = & \vec{\nabla}^{2}\vec{\psi}+\vec{\omega},\label{eq:psi}\end{eqnarray}
 where $\vec{g}=\vec{\nabla}\times(\vec{v}\cdot\vec{\nabla})\vec{v},$
and $\vec{h}=\vec{\nabla}\times\vec{f}$ are the \textit{curls} of
the dimensionless convective inertial and electromagnetic forces,
respectively and $\RE=\bar{w}_{0}d/\nu$ is the Reynolds number. In
fusion blanket applications $\RE\sim10^{1}-10^{5}$ while $\Ha\sim10^{3}-10^{4}.$

The boundary conditions for $\vec{\psi}$ and $\vec{\omega}$ are
obtained as follows. The impermeability condition applied integrally
as $\int_{s}\vec{v}\cdot\vec{ds}=\oint_{l}\vec{\psi}\cdot\vec{dl}=0$
to an arbitrary area of the wall $s$ encircled by a contour $l$
yields $\left.\psi_{\tau}\right|_{s}=0.$ Using this boundary condition,
which implies $\left.\vec{\psi}\right|_{s}=\left.\vec{n}\psi_{n}\right|_{s},$
in combination with (\ref{eq:divpsi}) we obtain $\left.\partial_{n}\psi_{n}\right|_{s}=0.$
In addition, the no-slip condition applied integrally $\oint_{l}\vec{v}\cdot\vec{dl}=\int_{s}\vec{\omega}\cdot\vec{ds}$
yields $\left.\omega_{n}\right|_{s}=0.$

We analyse linear stability of the base flow $\{\bar{\vec{\psi}},\bar{\vec{\omega}},\bar{\phi}\}(x,y)$
with respect to infinitesimal disturbances in the form of harmonic
waves travelling along the axis of the duct \[
\{\vec{\psi},\vec{\omega},\phi\}(\vec{r},t)=\{\bar{\vec{\psi}},\bar{\vec{\omega}},\bar{\phi}\}(x,y)+\{\hat{\vec{\psi}},\hat{\vec{\omega}},\hat{\phi}\}(x,y)e^{\gamma t+ikz},\]
 where $k$ is a wavenumber and $\gamma$ is, in general, a complex
growth rate. Upon substituting the solution sought in such a form
into (\ref{eq:omeg}), (\ref{eq:psi}) we obtain the governing equations
for the disturbance amplitudes \begin{eqnarray}
\gamma\hat{\vec{\omega}} & = & \vec{\nabla}_{k}^{2}\hat{\vec{\omega}}-\RE\hat{\vec{g}}+\Ha^{2}\hat{\vec{h}},\label{eq:omegh}\\
0 & = & \vec{\nabla}_{k}^{2}\hat{\vec{\psi}}+\hat{\vec{\omega}},\label{eq:psih}\\
0 & = & \vec{\nabla}_{k}^{2}\hat{\phi}-\hat{\omega}_{y},\label{eq:phih}\end{eqnarray}
 where $\vec{\nabla}_{k}\equiv\vec{\nabla}+ik\vec{e}_{z}.$ Because
of the solenoidity constraint satisfied by $\hat{\vec{\omega}}$ similarly
to $\hat{\vec{\psi}}$, we need only the $x$- and $y$-components
of (\ref{eq:omegh}), namely, $\hat{h}_{x}=-\partial_{xy}\hat{\phi}-\partial_{y}\hat{w},$
$\hat{h}_{y}=-\partial_{yy}\hat{\phi}$ and \begin{eqnarray}
\hat{g}_{x} & = & \quad k^{2}\hat{v}\bar{w}+\partial_{yy}(\hat{v}\bar{w})+\partial_{xy}(\hat{u}\bar{w})+i2k\partial_{y}(\hat{w}\bar{w}),\label{eq:gx}\\
\hat{g}_{y} & = & -k^{2}\hat{u}\bar{w}-\partial_{xx}(\hat{u}\bar{w})-\partial_{xy}(\hat{v}\bar{w})-i2k\partial_{x}(\hat{w}\bar{w}),\label{eq:gy}\end{eqnarray}
 where $\hat{u}=ik^{-1}(\partial_{yy}\hat{\psi}_{y}-k^{2}\hat{\psi}_{y}+\partial_{xy}\hat{\psi}_{x}),$
$\hat{v}=-ik^{-1}(\partial_{xx}\hat{\psi}_{x}-k^{2}\hat{\psi}_{x}+\partial_{xy}\hat{\psi}_{y}),$
and $\hat{w}=\partial_{x}\hat{\psi}_{y}-\partial_{y}\hat{\psi}_{x}.$
The relevant boundary conditions are\begin{eqnarray}
\partial_{x}\hat{\phi}=\hat{\psi}_{y}=\partial_{x}\hat{\psi}_{x}=\partial_{x}\hat{\psi}_{y}-\partial_{y}\hat{\psi}_{x}=\hat{\omega}_{x}=0 & \mbox{at} & x=\pm1,\label{eq:bndcx}\\
\hat{\phi}=\hat{\psi}_{x}=\partial_{y}\hat{\psi}_{y}=\partial_{x}\hat{\psi}_{y}-\partial_{y}\hat{\psi}_{x}=\hat{\omega}_{y}=0 & \mbox{at} & y=\pm A.\label{eq:bndcy}\end{eqnarray}

\section{Numerical method}

\begin{figure}
\begin{centering}
\includegraphics[width=0.75\textwidth]{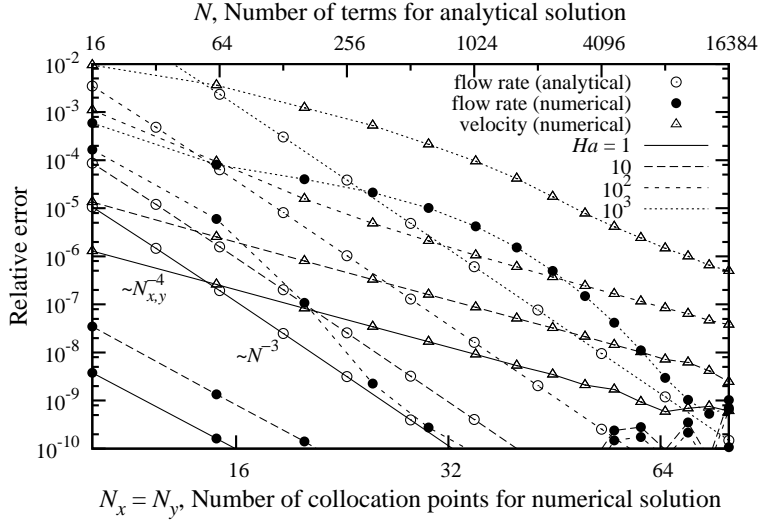} 
\par\end{centering}

\caption{\label{fig:hunt_err}Accuracy of the base flow at various Hartmann
numbers depending on the number of terms in the Fourier series solution
(top axis) and the number of collocation points for numerical solution
(bottom axis).}

\end{figure}

We solve the problem posed by (\ref{eq:wbar})-(\ref{eq:bbar}) and
(\ref{eq:omegh})-(\ref{eq:phih}) with the boundary conditions (\ref{eq:bndcx}),
(\ref{eq:bndcy}) by a spectral collocation method on a Chebyshev-Lobatto
grid using even number of points in the $x$- and $y$-directions
given by $2N_{x}+2$ and $2N_{y}+2,$ respectively, where $N_{x,y}=30\cdots55$
is used depending on $\Ha$ and $\RE.$ The convergence of the numerical
solution for the base flow was validated against the Fourier series
solution of \cite{Hun65}. First, we looked at the relative error
in the flow rate for a fixed pressure gradient. As seen in figure
\ref{fig:hunt_err}, analytical solution for this quantity converges
as $O(N^{-3})$ and requires $\approx10^{3}$ terms at $\Ha=10^{3}$
for the relative accuracy of $\approx10^{-6}.$ The numerical solution
for the flow rate shows a faster-than-algebraic convergence rate developing
at sufficiently high resolution, which is typical for spectral methods.
Second, maximum error in velocity scaled with respect to the velocity
maximum for $N_{x}=N_{y}$ decreases as $O(N_{x,y}^{-4}).$ For $\Ha=10^{3},$
the resolution of $N_{x}\times N_{y}=50\times50$ ensures the relative
accuracy in the base flow velocity of about $10^{-5}.$ The convergence
of the linear stability problem, for which the resolution of the base
flow is necessary but not sufficient, is tested below.

Because of the double reflection symmetry of the base flow with respect
to $x=0$ and $y=0$ planes, small-amplitude perturbations with different
parities in $x$ and $y$ decouple from each other. This results in
four mutually independent modes which we classify as $(o,o),$ $(o,e),$
$(e,o),$ and $(e,e)$ according to whether the $x$ and $y$ symmetry
of $\hat{\psi}_{x}$ is odd or even, respectively. Our classification
of modes specified in table \ref{tab:par} corresponds to the symmetries
I, II, III, and IV used by \cite{TatYos90} and \cite{UhlNag06}.
As a result, the problem is broken up into four independent problems
of different symmetries defined in one quadrant of the duct cross-section
with $N_{x}\times N_{y}$ internal collocation points. This allows
us to reduce the size of the matrix in the eigenvalue problem, which
is derived below, by a factor of 16. For each symmetry, we represent
(\ref{eq:omegh}), (\ref{eq:psih}) in the matrix form\begin{eqnarray}
\gamma\tilde{\omega}_{0} & = & \tilde{A}_{0}\tilde{\omega}_{0}+\tilde{A}_{1}\tilde{\omega}_{1}+\tilde{f}_{0},\label{eq:omeg0}\\
0 & = & \tilde{B}_{0}\tilde{\psi}_{0}+\tilde{\omega}_{0},\label{eq:psi0}\end{eqnarray}
 where $\tilde{\psi}_{0}$ and $\tilde{\omega}_{0}$ are the values
of $(\hat{\psi}_{x},\hat{\psi}_{y})$, $(\hat{\omega}_{x},\hat{\omega}_{y})$
at the internal collocation points, $\tilde{\omega}_{1}$ are unknown
values of the tangential component of $(\hat{\omega}_{x},\hat{\omega}_{y})$
at $N_{x}+N_{y}$ at boundary points; $\tilde{f}_{0}$ stands for
the source term in (\ref{eq:omegh}), $\tilde{A}_{0},$ $\tilde{A}_{1},$
and $\tilde{B}_{0}$ matrices represent collocation approximation
of  $\vec{\nabla}_{k}^{2}$ operator with the explicit boundary conditions
(\ref{eq:bndcx}), (\ref{eq:bndcy}) eliminated. For the unknown boundary
values of $\tilde{\omega}_{1},$ we have an extra boundary condition
$\partial_{x}\hat{\psi}_{y}-\partial_{y}\hat{\psi}_{x}=0$ imposed
on $(\hat{\psi}_{x},\hat{\psi}_{y})$ at $N_{x}+N_{y}$ boundary points
which is represented as \begin{equation}
\tilde{C}_{0}\tilde{\psi}_{0}=0.\label{eq:bndc0}\end{equation}
 To obtain a conventional matrix eigenvalue problem for $\gamma,$
we need to eliminate $\tilde{\omega}_{1}$ from (\ref{eq:omeg0}),
(\ref{eq:psi0}). Multiplying both sides of (\ref{eq:omeg0}) by $\tilde{C}_{0}\tilde{B}_{0}^{-1}$
we obtain \begin{equation}
\tilde{C}_{0}\tilde{B}_{0}^{-1}\tilde{A}_{1}\tilde{\omega}_{1}=-\tilde{C}_{0}\tilde{B}_{0}^{-1}(\tilde{A}_{0}\tilde{\omega}_{0}+\tilde{f}_{0}),\label{eq:omeg1}\end{equation}
 because (\ref{eq:psi0}), (\ref{eq:bndc0}) imply \begin{equation}
\gamma\tilde{C}_{0}\tilde{B}_{0}^{-1}\tilde{\omega}_{0}=-\gamma\tilde{C}_{0}\tilde{\psi}_{0}=0.\label{eq:bndc1}\end{equation}
 Now, $\tilde{\omega}_{1}$ can be expressed in terms of $\tilde{\omega}_{0}$
and $\tilde{f}_{0}$ by solving (\ref{eq:omeg1}) that substituted
back into (\ref{eq:omeg0}) results in \[
\gamma\tilde{\omega}_{0}=\tilde{D}_{0}(\tilde{A}_{0}\tilde{\omega}_{0}+\tilde{f}_{0}),\]
 where $\tilde{D}_{0}=\tilde{I}-\tilde{A}_{1}(\tilde{C}_{0}\tilde{B}_{0}^{-1}\tilde{A}_{1})^{-1}\tilde{C}_{0}\tilde{B}_{0}^{-1}$
and $\tilde{I}$ is the identity matrix. Note that $\tilde{f}_{0}$
is linear in both $\tilde{\psi}_{0}$ and $\tilde{\phi}_{0}$ where
the latter can be expressed as $\tilde{\phi}_{0}=\tilde{E}_{0}^{-1}\tilde{\omega}_{0,y}$
by solving the matrix counterpart of (\ref{eq:phih}). Eventually,
using (\ref{eq:psi0}), we can write $\tilde{f}_{0}=\tilde{F}_{0}\tilde{\psi}_{0},$
that leads to \begin{equation}
\gamma\tilde{\psi}_{0}=\tilde{B}_{0}^{-1}\tilde{D}_{0}(\tilde{A}_{0}\tilde{B}_{0}-\tilde{F}_{0})\tilde{\psi}_{0}.\label{eq:eigv0}\end{equation}
 This complex matrix eigenvalue problem is solved by the LAPACK's
ZGEEV routine.

\begin{table}
\begin{centering}
\begin{tabular}{rcccc}
 & I & II & III & IV\tabularnewline
$\hat{\psi}_{x},\hat{\omega}_{x},\hat{v}:$ & (o,o) & (o,e) & (e,o) & (e,e)\tabularnewline
$\hat{w}:$ & (o,e) & (o,o) & (e,e) & (e,o)\tabularnewline
$\hat{\psi}_{z},\hat{\omega}_{z}:$ & (e,o) & (e,e) & (o,o) & (o,e)\tabularnewline
$\hat{\psi}_{y},\hat{\omega}_{y},\hat{u},\hat{\phi}:$ & (e,e) & (e,o) & (o,e) & (o,o)\tabularnewline
\end{tabular}
\par\end{centering}

\caption{\label{tab:par}The $(x,y)$-parities of different variables for symmetries
I, II, III and IV; $e$ - even, $o$ - odd. }

\end{table}

\begin{table}
\begin{centering}
\begin{tabular}{cccc}
$N_{x}\times N_{y}$  & $c$ & $N_{x}\times N_{y}$ & $c$\tabularnewline
$20\times20$  & $(0.9515252,-0.02267611)$ & $20\times35$ & $(0.23219007,-0.3204544\times10^{-3})$\tabularnewline
$24\times24$  & $(0.9514767,-0.02258387)$ & $25\times40$ & $(0.23237696,-0.5640044\times10^{-4})$\tabularnewline
$28\times28$  & $(0.9514760,-0.02258432)$ & $30\times45$ & $(0.23239397,-0.28703331\times10^{-4})$\tabularnewline
$32\times32$  & $(0.9514760,-0.02258432)$ & $30\times50$ & $(0.23239343,-0.21204144\times10^{-4})$\tabularnewline
$36\times36$  & $(0.9514760,-0.02258432)$ & $40\times60$ & $(0.23239274,-0.21981102\times10^{-4})$\tabularnewline
\end{tabular}
\par\end{centering}

\caption{\label{tab:val}Convergence of the complex relative phase velocity
$c=i\gamma/(\RE k)$ of the least stable mode of symmetry I for a
model base flow $\bar{w}(x,y)=(1-x^{2})(A^{2}-y^{2})$ with $A=1,$
$\RE=10^{4}$ and $k=1$ (left) and for the non-magnetic duct flow
with $A=5,$ $\RE=1.04\times10^{4}$ and $k=0.91$ considered by \cite{TatYos90}
(right). }

\end{table}

Without the base flow $(\RE=0),$ the leading eigenvalues of (\ref{eq:eigv0})
are real and negative except for $N_{x}+N_{y}$ eigenvalues which
are zero within machine accuracy. These spurious eigenvalues are caused
by the way the boundary condition (\ref{eq:bndc0}) is imposed using
(\ref{eq:bndc1}) which can also be satisfied by $\gamma=0.$ These
zero eigenvalues can easily be identified and discarded. Alternatively,
they can be shifted down the spectrum by an arbitrary value $\gamma_{0}$
when $\gamma_{0}\tilde{C}_{0}\tilde{\psi}_{0}$ is added to the right-hand
side of (\ref{eq:omeg1}). Note that this transformation does not
affect the true eigenmodes which satisfy the boundary condition (\ref{eq:bndc0}).
However, our approach is not completely free of unstable spurious
eigenmodes which may appear at sufficiently high $\RE$ depending
on the collocation approximation of inertial terms (\ref{eq:gx}),
(\ref{eq:gy}). Because the collocation differentiation satisfy the
product rule approximately rather than exactly (\cite{Fron96}), the
discretisation of inertial terms is affected by the form in which
they are presented. We find that the number of unstable spurious eigenmodes
is the least when the inertial terms are approximated in the {}``conservative''
form given by (\ref{eq:gx}), (\ref{eq:gy}). In contrast to the true
eigenmodes, the spurious ones are numerical artifacts which depend
strongly on the number of collocation points. This allows us to identify
them easily by recalculating the spectrum with $N_{x}+1$ and $N_{y}+1$
collocation points and retaining only those eigenvalues whose modulus
of the relative variation is typically less than $\varepsilon=10^{-3}-10^{-4},$
which is subsequently referred to as the relative accuracy threshold.
Once a true eigenvalue is identified, it can be tracked further by
its imaginary part without recalculating the spectrum as the control
parameters are slowly varied.

\begin{table}
\begin{centering}
\begin{tabular}{cccc}
$A$ & $\RE_{c}\times10^{-4}$ & $k_{c}$ & $c_{c}$\tabularnewline
5 & $1.043$ & $0.9085$ & $0.2321$\tabularnewline
4 & $1.819$ & $0.8139$ & $0.2042$\tabularnewline
3.5 & $3.650$ & $0.7075$ & $0.1738$\tabularnewline
\end{tabular}
\par\end{centering}

\caption{\label{tab:nonm}The critical Reynolds number $\RE_{c}$, wavenumber
$k_{c}$ and phase velocity $c_{c}$ obtained with $30\times50$ collocation
points for the instability mode I in the non-magnetic duct flow at
various aspect ratios $A.$}

\end{table}

\begin{table}
\begin{centering}
\begin{tabular}{cc}
$N_{x}\times N_{y}$  & $c$~$(\Ha=10)$\tabularnewline
$20\times20$  & $(0.7579394,-0.3312792\times10^{-3})$\tabularnewline
$25\times25$  & $(0.7579419,-0.3334821\times10^{-3})$\tabularnewline
$30\times30$  & $(0.7579419,-0.3337441\times10^{-3})$\tabularnewline
$35\times35$  & $(0.7579413,-0.3337346\times10^{-3})$\tabularnewline
$40\times40$  & $(0.7579413,-0.3337034\times10^{-3})$\tabularnewline
\end{tabular} \begin{tabular}{cc}
$N_{x}\times N_{y}$  & $c$~$(\Ha=10^{2})$\tabularnewline
$25\times25$  & $(0.4907420,-0.8028854\times10^{-2})$\tabularnewline
$30\times30$  & $(0.4907416,-0.8028572\times10^{-2})$\tabularnewline
$35\times35$  & $(0.4907415,-0.8028551\times10^{-2})$\tabularnewline
$40\times40$  & $(0.4907415,-0.802855\times10^{-2})$\tabularnewline
$45\times45$ & $(0.4907415,-0.802855\times10^{-2})$\tabularnewline
\end{tabular}
\par\end{centering}

\begin{centering}
\begin{tabular}{cc}
$N_{x}\times N_{y}$  & $c$~$(\Ha=10^{3})$\tabularnewline
$40\times40$  & $(0.5053194,0.1453188\times10^{-2})$\tabularnewline
$45\times45$  & $(0.5054035,0.1421803\times10^{-2})$\tabularnewline
$50\times50$  & $(0.5053892,0.1416928\times10^{-2})$\tabularnewline
$55\times55$  & $(0.5053904,0.1416891\times10^{-2})$\tabularnewline
$60\times60$ & $(0.5053902,0.1417051\times10^{-2})$\tabularnewline
\end{tabular}
\par\end{centering}

\caption{\label{tab:Hunt}Convergence of the complex relative phase velocity
$c=i\gamma/(\RE k)$ of the least stable mode for the Hunt's flow
in square duct at three different Hartmann numbers: 1) $\Ha=10,$
$\RE=2000,$ $k=0.8$, mode II; 2) $\Ha=10^{2},$ $\RE=10^{3},$ $k=5,$
mode I; 3) $\Ha=10^{3},$ $\RE=3\times10^{3},$ $k=16,$ mode I. }

\end{table}

The numerical method has been validated using a model base flow $\bar{w}(x,y)=(1-x^{2})(A^{2}-y^{2})$
with $A=1$ at $\RE=10^{4}$ and $k=1$ as well as the non-magnetic
duct flow with $A=5,$ $\RE=1.04\times10^{5}$ and $k=0.91$ considered
by \cite{TatYos90} that resulted in the complex relative phase velocity
$c=i\gamma/(\RE k)$ for the least stable mode of symmetry I $(o,o)$
which is shown in table \ref{tab:val} for various resolutions. For
the model flow, the increase of the resolution from $20\times20$
to $28\times28$ collocation points results in the fast convergence
of the leading eigenvalue with the accuracy raising from two to seven
figures, respectively, which is comparable to the accuracy of the
Galerkin method for this test problem used by \cite{Uhl04}. Similarly
fast convergence is obvious also for the non-magnetic duct flow with
aspect ratio $A=5.$ Owing to the large aspect ratio $A=5$ as well
as the high Reynolds number $\RE=1.04\times10^{4},$ which for $k=0.91$
is close to its critical value, at least $30\times45$ collocation
points are required to obtain the phase velocity with 5 accurate figures,
which again is comparable to the accuracy of the Galerkin method tested
against the same case by \cite{UhlNag06}. Also the instability threshold
parameters for the aspect ratios $A=5,4,3.5,$ which are shown in
table \ref{tab:nonm} for $30\times50$ resolution, agree well with
\cite{TatYos90}.

\begin{figure}
\begin{centering}
\includegraphics[width=0.75\textwidth]{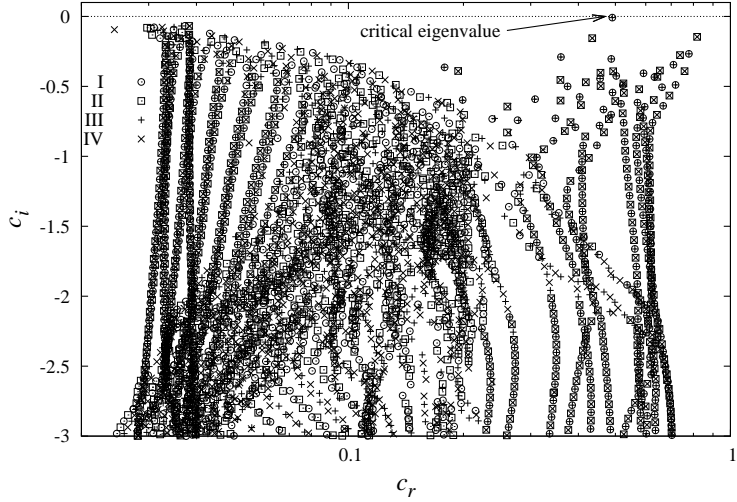} 
\par\end{centering}

\caption{\label{fig:legv}Spectrum of the complex relative phase velocities
$c=i\gamma/(\RE k)$ for all four mode types at $\Ha=100,$ $\RE=10^{3},$
$k=5$ obtained with $50\times50$ collocation points and the relative
accuracy threshold $\varepsilon=10^{-3}$.}

\end{figure}

As seen in table \ref{tab:Hunt}, a comparably fast convergence holds
also for the complex phase velocity of the least stable modes in the
Hunt's flow at $\Ha=10,10^{2},10^{3}.$ Detailed numerical results
for these instability modes are presented in the next section. A typical
spectrum of the complex relative phase velocities $c$ is shown in
figure \ref{fig:legv} for $\Ha=100$ close to the instability threshold
for the least stable modes of type I and III. The eigenvalues have
been computed using $50\times50$ collocation points and the relative
accuracy threshold $\varepsilon=10^{-3}.$

Subsequently, to verify the numerical accuracy of the obtained results
we recalculate them with the resolution increased by 5 collocation
points in each direction. Only the results coinciding by at least
four leading figures are retained. For modes I and III, the resolution
of $35\times35$ ensures the accuracy of at least $5$ digits at $\Ha=100,$
while $50\times50$ resolution is required at $\Ha=3\times10^{3}.$
Modes II and IV require only $30\times30$ resolution at $\Ha\approx10,$
whereas $55\times55$ points are required at $\Ha\approx400.$


%

\section{Results and discussion}

Here we present the results for the flow in a square duct which according
to \cite{TatYos90} is linearly stable in the non-magnetic case. First,
we find the base flow numerically and normalise it with respect to
its maximum velocity which is used here as the velocity scale. The
flow rate over one quarter of the duct, which is also the average
velocity, is found to vary for $\Ha\gg1$ as \begin{equation}
Q\approx1.23\Ha^{-1/2}+3.87\Ha^{-1},\label{eq:flwr}\end{equation}
 where both coefficients are obtained by the best fit of the numerical
solution. The main contribution to the flow rate is due to the side
jets whereas the next-order correction is due to the core flow. Although
the characteristic velocity of the core flow is only $O(\Ha^{-1})$
with respect to that of the side layers, its relative contribution
to the flow rate is $\Ha^{1/2}$ times larger because the relative
thickness of side jets is $O(\Ha^{-1/2}).$ If the flow rate were
used for the characteristic velocity, the relative contribution of
the core flow in the critical Reynolds number would be $O(\Ha^{-1/2}).$
In contrast, when the maximum velocity is used for this purpose, the
correction is only $O(\Ha^{-1})$ which becomes negligible at a much
lower $\Ha$ than the previous one. This results in a more definite
asymptotics appearing at numerically attainable values of $\Ha.$
Note that for $\Ha=10^{3}$ the relative contribution of the core
flow to the flow rate is about $10\%.$ Therefore, we have chosen
the maximum rather than average velocity as the characteristic scale.

\begin{figure}
\begin{centering}
\includegraphics[width=0.5\textwidth]{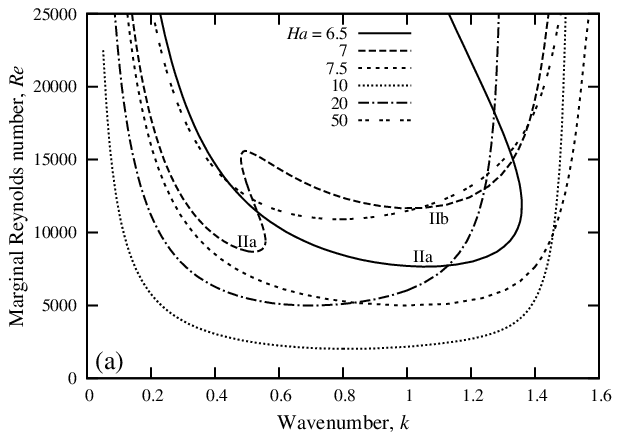}\includegraphics[width=0.5\textwidth]{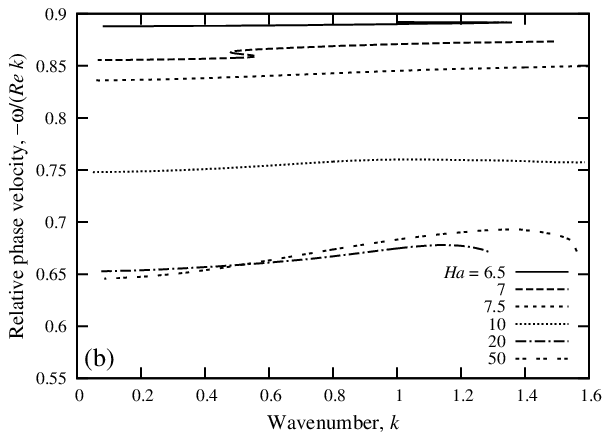} 
\par\end{centering}

\caption{\label{fig:rewk_Ge1_12c}Marginal Reynolds number (a) and relative
phase velocity (b) versus the wavenumber for neutrally stable modes
of type II. }

\end{figure}

\begin{figure}
\begin{centering}
\includegraphics[width=0.5\textwidth]{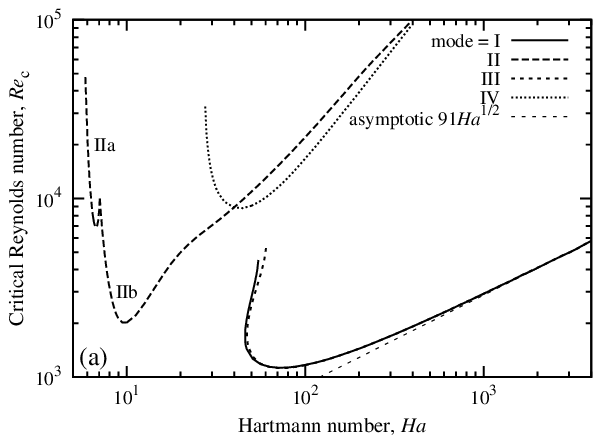}\includegraphics[width=0.5\textwidth]{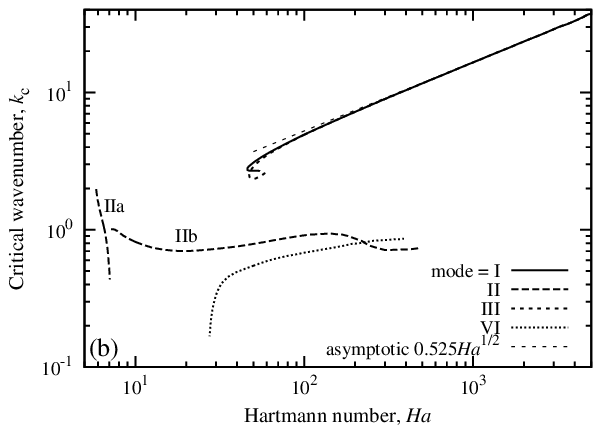}\\
 \includegraphics[width=0.5\textwidth]{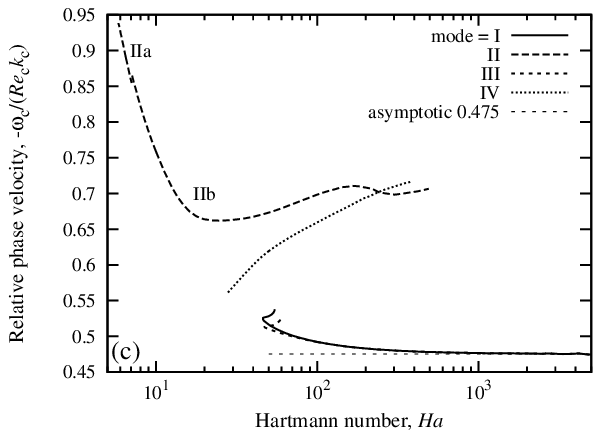} 
\par\end{centering}

\caption{\label{fig:crt_Ha} Critical Reynolds number (a), wavenumber (b) and
relative phase velocity (c) against Hartmann number.}

\end{figure}

The neutral stability curves for the instability type II plotted in
figure \ref{fig:rewk_Ge1_12c} show the marginal Reynolds number,
which yields zero growth rate $(\Re[\gamma]=0)$ of the most unstable
mode for the given wavenumber, and the relative phase velocity $c=-\omega/(\RE k)$
at various Hartmann numbers. Here $\omega=\Im[\gamma]$ is the frequency
of the corresponding neutrally stable mode. The minimum of the marginal
Reynolds number and the corresponding wavenumber at which it occurs
give, respectively, the critical value $\RE_{c}$ and the critical
wavenumber $k_{c}.$ This wavenumber along with the corresponding
phase velocity is plotted in figure \ref{fig:crt_Ha} against the
Hartmann number. It is seen in figure \ref{fig:crt_Ha}(a) that the
mode of type II, which is the most unstable up to $\Ha\approx40,$
first appears at $\Ha\approx5.7$. At this Hartmann number, the velocity
profile of the base flow, which is very close to those shown in figures
\ref{fig:sketch}(b) and (c) for $\Ha=6,$ has a minimum at the centre
of the duct accompanied by two slight maxima at the each side of it.
With the increase of the magnetic field, these velocity maxima develop
into the jets localised at the side walls of the duct (see figure
\ref{fig:sketch}c). There are inflection points in the velocity profile,
which imply a possibility of an inviscid-type instability, however
this criterion is generally restricted to one-dimensional inviscid
flows (\cite{BOH88}).

Note that in a certain range of the Hartmann number there may be two
local minima on the neutral stability curve. These are denoted as
(IIa) and (IIb) in figure \ref{fig:rewk_Ge1_12c}(a). The first minimum,
IIa, is below the second one up to $\Ha\approx7$ where the critical
mode mode switches to IIb. The corresponding branches of the critical
parameters for this mode are labelled as IIa and IIb in figure \ref{fig:crt_Ha}.
With the increase of $\Ha$, $\RE_{c}$ first steeply decreases down
to its minimal value of $\RE_{c}\approx2018$ at $\Ha\approx10$ and
then starts to increase with the rate becoming nearly proportional
to $\Ha$ for $\Ha\gtrsim40.$ It is important to note that the relative
phase velocity of the neutrally stable modes, shown in figure \ref{fig:rewk_Ge1_12c}(b),
is nearly invariant with wavenumber, and has the order of magnitude
$O(1).$ Moreover, the relative phase velocity is seen in figure \ref{fig:crt_Ha}(c)
to stay about $O(1)$ at large $\Ha$ as well. Both of these facts
imply that the phase velocity of unstable modes is strongly correlated
with the maximum velocity defined by $\RE.$

In order to visualise the three-dimensional velocity field of the
critical perturbation given by $\Re[\hat{\vec{v}}(x,y)e^{ik_{c}z}]$
we consider the complex amplitude of the velocity perturbation $\hat{\vec{v}}=\vec{\nabla}_{k}\times\hat{\vec{\psi}}$
which is associated with the corresponding vector stream function
$\hat{\vec{\psi}}.$ The velocity field $(\hat{u},\hat{v})$ in the
$(x,y)$-plane can be decomposed into solenoidal $\hat{\vec{v}}_{s}$
and potential $\hat{\vec{v}}_{p}$ components which satisfy $\vec{\nabla}\cdot\hat{\vec{v}}_{s}=0$
and $\vec{\nabla}\times\hat{\vec{v}}_{p}=0,$ respectively. The solenoidal
component satisfying the impermeability boundary condition is given
by $\hat{\vec{v}}_{s}=-\vec{e}_{z}\times\vec{\nabla}\hat{\psi}_{z}$
which implies that $\hat{\psi}_{z}$ is the stream function of $\hat{\vec{v}}_{s}.$
The incompressibility constraint of the whole velocity perturbation,
which may be written as $\vec{\nabla}\cdot\hat{\vec{v}}_{p}=-ik\hat{w},$
in turn, links the potential component $\hat{\vec{v}}_{p}$ to the
longitudinal velocity perturbation $\hat{w},$ which serves as a source
or a sink for the former. Therefore, the whole velocity perturbation
is completely defined by $\hat{\psi}_{z}$ and $\hat{w}.$  In a similar
way, the streamlines of solenoidal flow components in the $(x,z)$-
and $(y,z)$-planes are given by $\hat{\psi}_{y}$ and $\hat{\psi}_{x},$
respectively. Note that the perturbation amplitudes are complex quantities
whose real and imaginary parts correspond to the instantaneous distributions
in the $(x,y)$-plane shifted in time or in space by a quarter of
a period.

\begin{figure}
\begin{centering}
\includegraphics[bb=150bp 95bp 320bp 255bp,clip,width=0.5\textwidth]{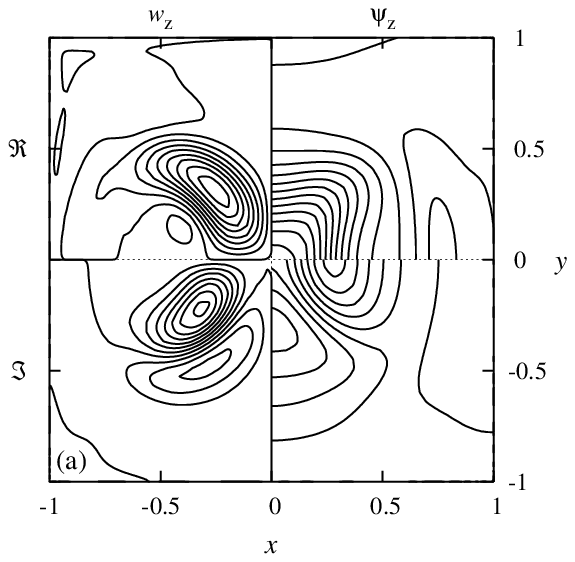}\includegraphics[bb=150bp 95bp 320bp 255bp,clip,width=0.5\textwidth]{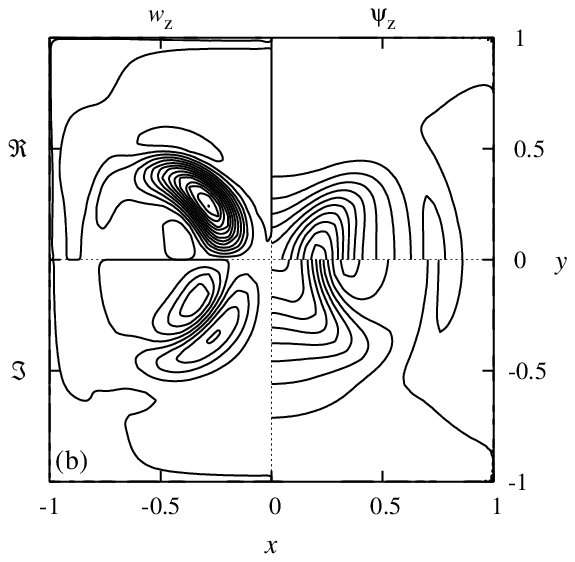}
\par\end{centering}

\caption{\label{fig:egv_Ge1_Ha7v_12} Amplitude distributions of real $(y>0)$
and imaginary $(y<0)$ parts of $\hat{w}$ $(x<0)$ and $\hat{\psi}_{z}$
$(x>0)$ of the critical perturbations over one quadrant of duct cross-section
for instability modes IIa (a) and IIb (b) at $\Ha\approx7$ and $\RE\approx10^{4}.$}

\end{figure}

Distributions of the most unstable perturbation amplitudes of types
IIa and IIb are plotted in figure \ref{fig:egv_Ge1_Ha7v_12} for $\Ha\approx7$
over different quadrants of the duct cross-section. Both perturbations
differ mainly by the critical wavenumbers, $k_{c}\approx0.44$ and
$k_{c}\approx1,$ respectively, but have similar amplitude distributions
concentrated about the centre of the duct. Transversal circulation
in the $(x,y)$-plane, which is given by the isolines of $\hat{\psi}_{z},$
takes place about the centre of the duct with $\hat{v}_{x}$ and $\hat{v}_{y}$
being even functions of $x$ and $y,$ respectively. The longitudinal
velocity perturbation $\hat{w},$ caused by the advection of momentum
of the base flow by the transversal circulation, is an odd function
of both $x$ and $y.$ Both of these instability modes are obviously
related to the two local velocity maxima which appear first in the
centre of the base flow at $\Ha\approx6.$ With the increase of $\Ha$
these two velocity maxima develop into a pair of jets along the insulating
side walls (see figure \ref{fig:sketch}b, $x<0$).

\begin{figure}
\begin{centering}
\includegraphics[width=0.5\textwidth]{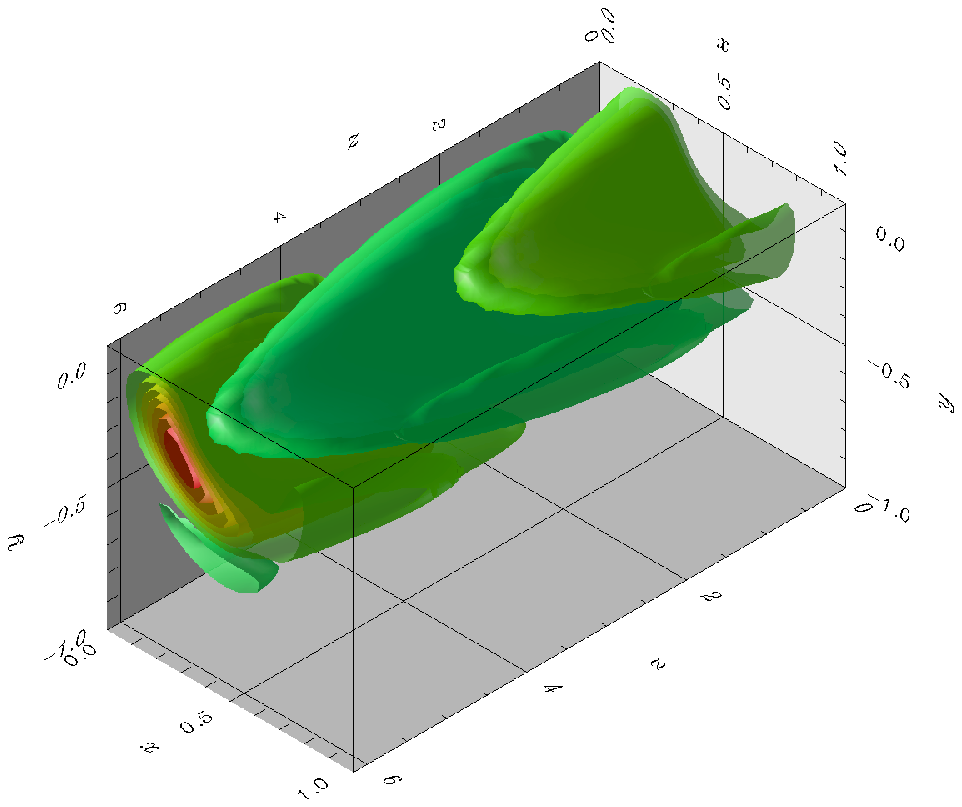}\put(-100,10){(a)}\includegraphics[width=0.5\textwidth]{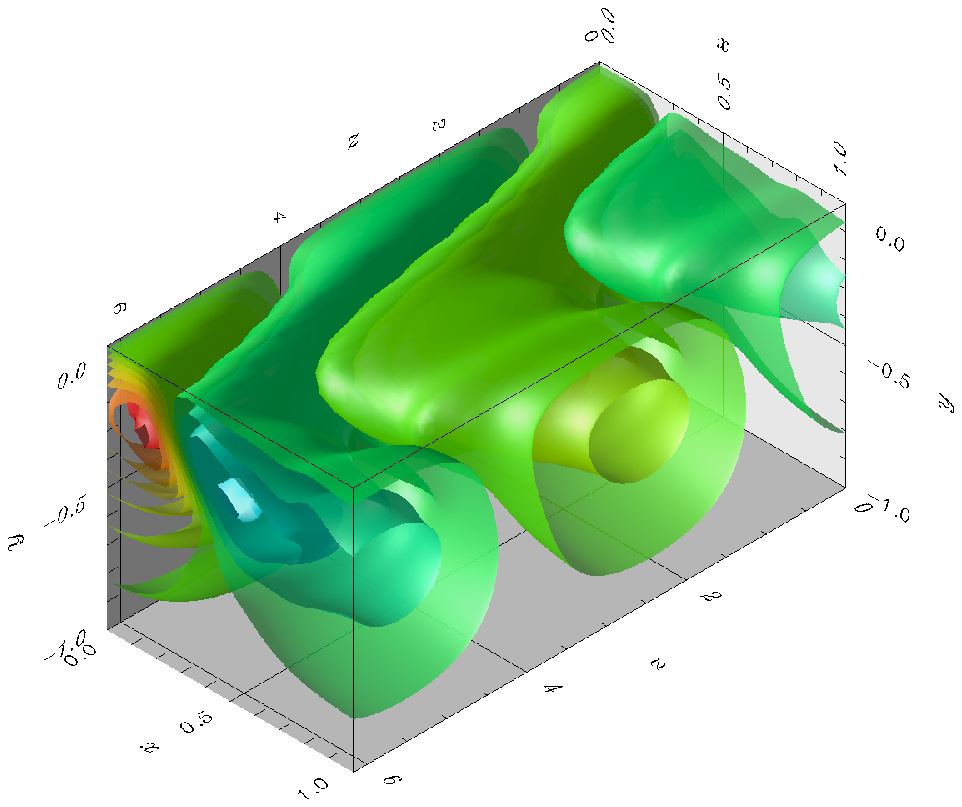}\put(-100,10){(b)}
\par\end{centering}

\caption{\label{fig:wz_Ge1_Ha7v_12} Isosurfaces of longitudinal velocity $\hat{w}$
(a) and electric potential $\hat{\phi}$ (b) perturbation over a wavelength
in one quadrant of duct cross-section for instability mode IIb at
$\Ha\approx7$ and $\RE\approx10^{4}.$ }

\end{figure}

Besides spatial amplitude distributions, the perturbations can be
characterised by the kinetic energy distribution over the velocity
or vorticity/stream function components as follows: \[
E\propto\int_{S}\hat{\left|\vec{v}\right|}^{2}\, ds=\int_{S}\Re[\hat{\vec{\omega}}\cdot\hat{\vec{\psi}}^{*}]\, ds,\]
 where $E$ is the kinetic energy of perturbation averaged over the
wavelength. The integrals in the expression above are taken over the
duct cross-section $S,$ and the asterisk denotes the complex conjugate.
We find that $98\%$ and $91\%$ of kinetic energy for modes IIa and
IIb, respectively, are carried by the longitudinal velocity perturbation
$\hat{w}.$ The corresponding component of the vorticity perturbation
$(\hat{\omega}_{z},\hat{\psi}_{z}),$ which is associated with the
circulation in the $(x,y)$-plane, contains only $2\%$ and $7\%$
of kinetic energy, respectively. The isosurfaces of the critical perturbations
of longitudinal velocity and electric potential are shown in figure
\ref{fig:wz_Ge1_Ha7v_12} for one wavelength of mode IIb in the right
bottom quadrant of duct at $\Ha=7$ and $\RE=10^{4}.$ The corresponding
perturbation pattern for mode IIa differs mainly by a longer wavelength.
As seen in figure \ref{fig:wz_Ge1_Ha7v_12}(a), the perturbation of
$\hat{w}$ represents a pair of elongated, slightly tilted and periodically
overlapping streaks located close to the centre of the duct. The perturbation
of the electric potential, which is the largest in the vertical mid-plane
of the duct $(x=0)$, partly reaches the side walls where it can be
measured experimentally.

\begin{figure}
\begin{centering}
\includegraphics[width=0.5\textwidth]{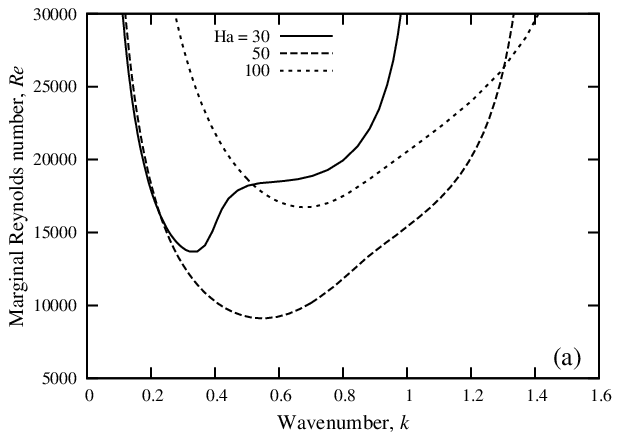}\includegraphics[width=0.5\textwidth]{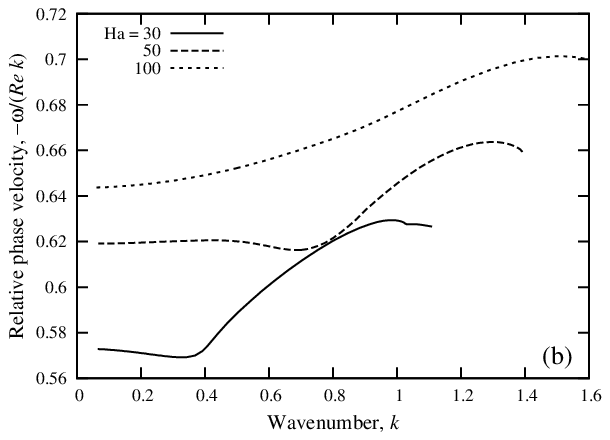} 
\par\end{centering}

\caption{\label{fig:rewk_Ge1_22c}Marginal Reynolds number (a) and relative
phase velocity (b) versus the wavenumber for neutrally stable modes
of type IV.}

\end{figure}

Neutral stability curves for the instability mode of type IV, which
appears for $\Ha\gtrsim28$ and differs from the previous one by the
opposite $x$-parity, are plotted in figure \ref{fig:rewk_Ge1_22c}.
Figure \ref{fig:crt_Ha} shows that $\RE_{c}$ of this mode, which for
low values of $\Ha$ lies above that of mode II, first steeply decreases
with $\Ha$ by reaching $\RE_{c}\approx9\times10^{3}$ of mode II
at $\Ha\approx40.$ The critical Reynolds number for mode IV attains
a minimum of $\RE_{c}\approx8.8\times10^{3}$ at $\Ha\approx44$ and
then starts to increase with $\Ha$ remaining below $\RE_{c}$ for
mode II up to the largest numerically attainable value of $\Ha\approx400.$

\begin{figure}
\begin{centering}
\includegraphics[bb=150bp 95bp 320bp 255bp,clip,width=0.5\textwidth]{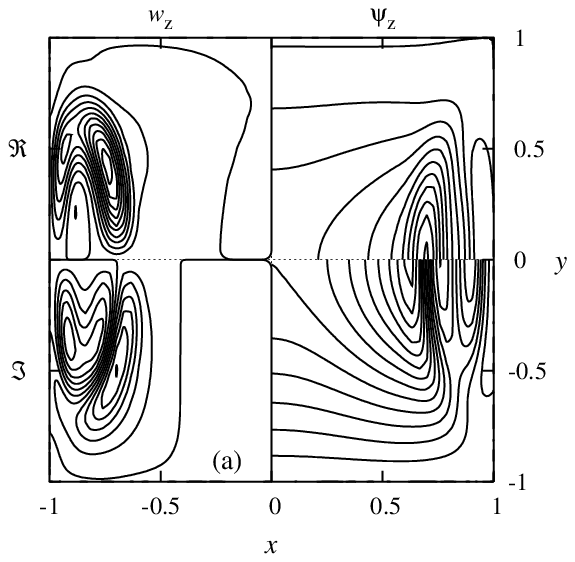}\includegraphics[bb=150bp 95bp 320bp 255bp,clip,width=0.5\textwidth]{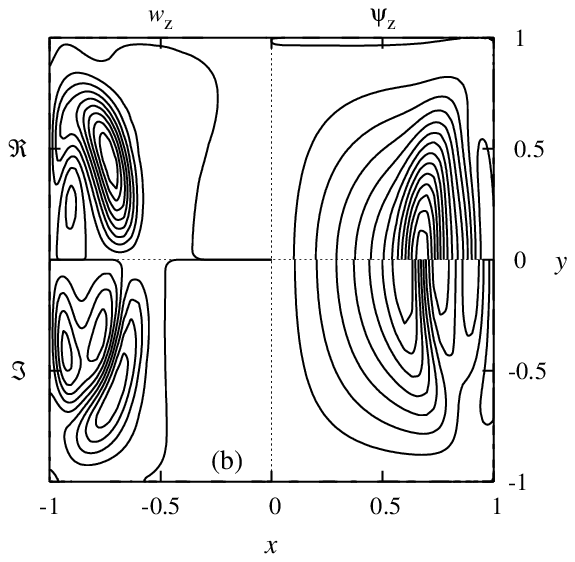}
\par\end{centering}

\caption{\label{fig:egv_Ge1_Ha40v_12-22} Amplitude distributions of real $(y>0)$
and imaginary $(y<0)$ parts of $\hat{w}$ $(x<0)$ and $\hat{\psi}_{z}$
$(x>0)$ of the critical perturbations over one quadrant of duct cross-section
for instability modes II (a) and IV (b) at $\Ha\approx40$ and $\RE\approx9\times10^{4}.$}

\end{figure}

Amplitude distributions of the most unstable perturbations of types
II and IV are plotted in figure \ref{fig:egv_Ge1_Ha40v_12-22} at
$\Ha\approx40$ and $\RE\approx9\times10^{3}.$ The critical wavenumbers
for these modes are, respectively, $k_{c}\approx0.75$ and $k_{c}\approx0.49.$
It is seen in figure \ref{fig:egv_Ge1_Ha40v_12-22}(a), that mode
II has moved from the centre of duct, where it originally appeared
at $\Ha\approx5.7$, to the side wall. The only principal difference
between these modes is the opposite $x$-parity which results in a
pair of mirror-symmetric longitudinal vortices on each side of the
duct with the same or opposite sense of circulation for modes II and
IV, respectively. In the first case, both vortices are partly connected
across the vertical mid-plane of the duct whereas they are separated
by that plane in the second case. For both modes, the perturbations
of the longitudinal velocity are localised in the sidewall jets and
are very similar to each other except for the opposite phases of oscillations
across the width of the duct. Modes II and IV are also similar from
the energetic point of view with $88\%$ and $93\%$ of kinetic energy
concentrated in the perturbation of the longitudinal velocity. The
least amount of energy, which is about $1\%$ and $0.4\%,$ respectively,
is contained in the $x$-component of the velocity perturbation while
the rest carried by the $y$-component parallel to the magnetic field.

\begin{figure}
\begin{centering}
\includegraphics[width=0.5\textwidth]{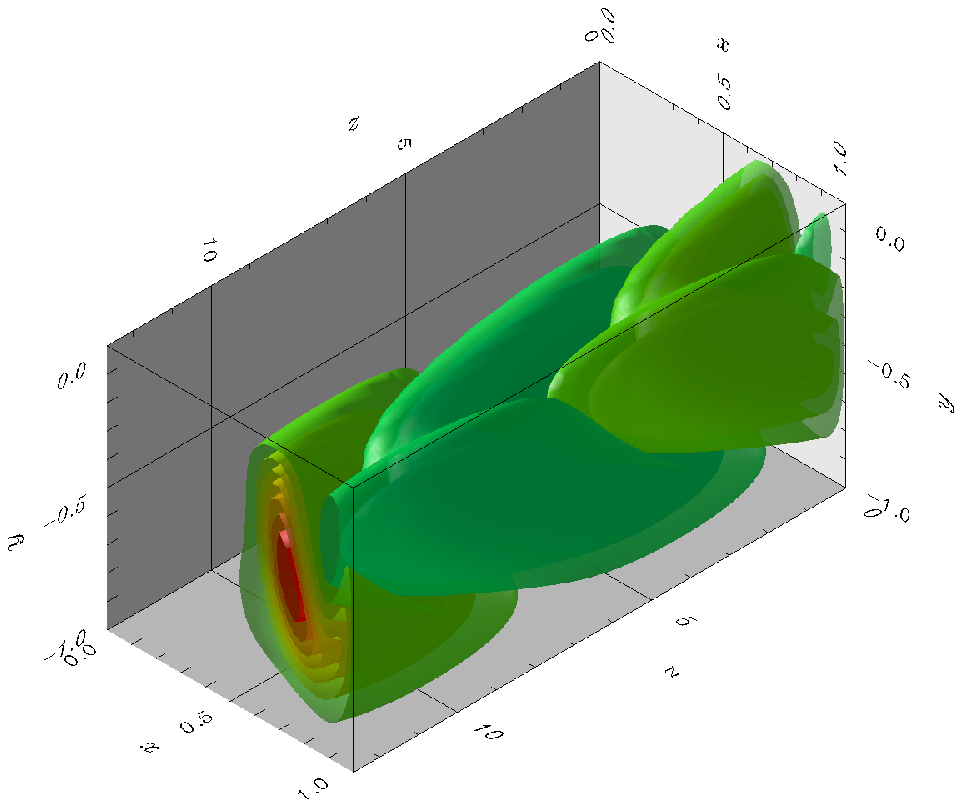}\put(-100,10){(a)}\includegraphics[width=0.5\textwidth]{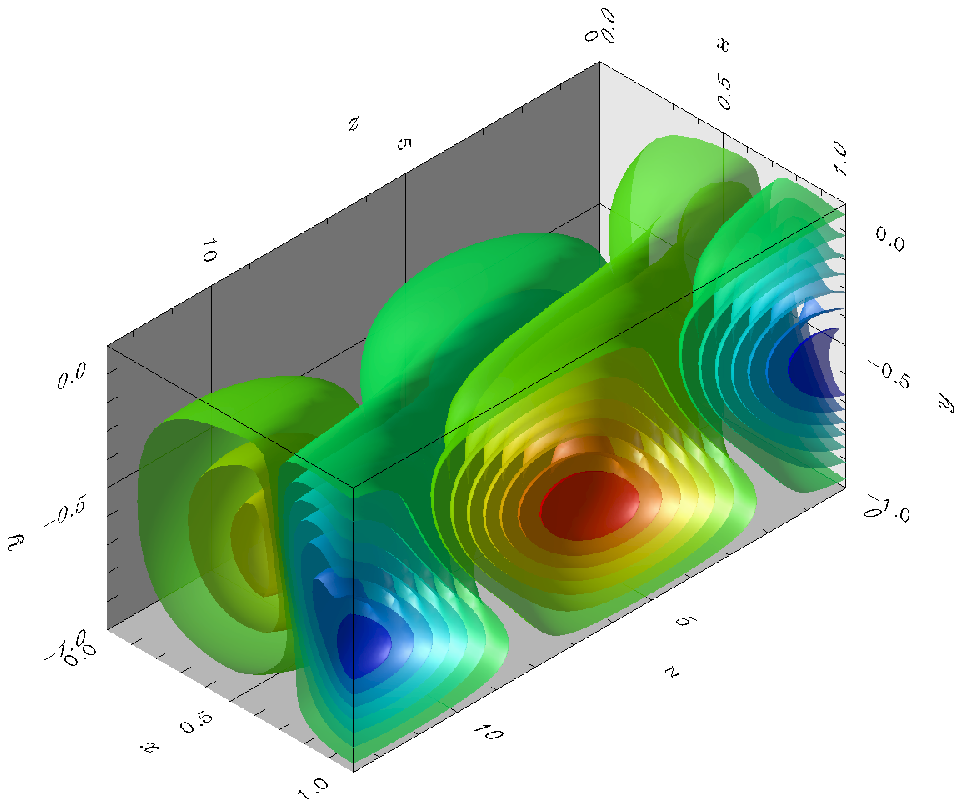}\put(-100,10){(b)}
\par\end{centering}

\caption{\label{fig:wz_Ge1_Ha40v_22} Isosurfaces of longitudinal velocity
$w$ (a) and electric potential $\phi$ (b) perturbation over a wavelength
in one quadrant of duct cross-section for instability mode IV at $\Ha\approx40$
and $\RE\approx9\times10^{4}.$}

\end{figure}

The isosurfaces of the critical perturbations of the longitudinal
velocity and of the electric potential for mode IV are shown in figure
\ref{fig:wz_Ge1_Ha40v_22} over one wavelength in the right bottom
quadrant of duct for $\Ha=40$ and $\RE\approx9\times10^{3}.$ In
this case, the distributions of $\hat{w}$ and $\hat{\phi}$ are even
and odd functions of $x,$ respectively. The corresponding pattern
for mode II at these parameters differs from that of mode IV mainly
by the shorter wavelength and opposite $x$-parity that results in
a non-zero perturbation of $\hat{\phi}$ in the vertical mid-plane
of the duct $(x=0)$.

\begin{figure}
\begin{centering}
\includegraphics[width=0.5\textwidth]{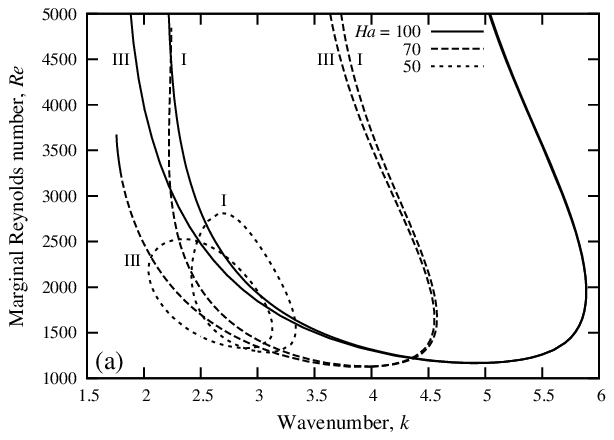}\includegraphics[width=0.5\textwidth]{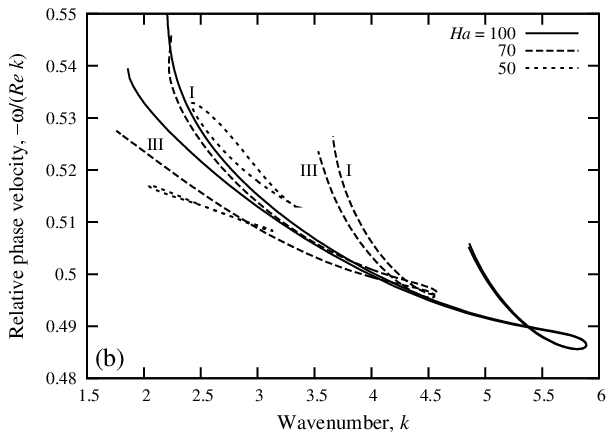} 
\par\end{centering}

\caption{\label{fig:rewk_Ge1_11-12c}Marginal Reynolds number (a) and relative
phase velocity (b) versus the wavenumber for neutrally stable modes
of types I and III. }

\end{figure}

A pair of additional instability modes of type I and III appears for
$\Ha\gtrsim46$ and $47,$ respectively. These modes differ from the
ones of type II and IV by the opposite $y$-parity. The neutral stability
curves plotted in figure \ref{fig:rewk_Ge1_11-12c} look very similar
for both of these modes. First, for $\Ha\lesssim54,\,60$ the neutral
stability curves are seen to form closed loops which implies that
both modes are unstable only within limited ranges of Reynolds and
wavenumbers.  In this range of $\Ha,$ there is not only the lower
but also the upper critical value of $\RE,$ by exceeding which all
perturbations of the corresponding type become linearly stable again.

These critical values of $\RE,$ which are considerably lower than
those for the previous two modes, are plotted in figure \ref{fig:crt_Ha}
against the Hartmann number along with the corresponding wavenumbers
and the relative phase velocities. As seen in figure \ref{fig:crt_Ha}(a),
the upper critical Reynolds number steeply increases with $\Ha$ becoming
very large at $\Ha\approx54,\,60$ for modes I and III, respectively.
The lower value of $\RE_{c}$ steeply decreases to its minimum $\RE_{c}\approx1130$
attained at $\Ha\approx70.$ A further increase of $\Ha$ results
in the growth of the critical Reynolds number for both modes approaching
the asymptotics $\RE_{c}\approx91\Ha^{1/2}$ for $\Ha\gg1.$ This
implies that the critical Reynolds number based on the average velocity
tends to a constant $\bar{\RE}_{c}\approx112$ while the next-order-correction
is about $352\Ha^{-1/2}.$ In contrast to this, the relative next-order-correction
for $\RE_{c}$ based on the maximal velocity, as discussed at the
beginning of this section, is only $O(\Ha^{-1}).$ The critical wavenumber
for both modes I and II tends to $k_{c}\approx0.525\Ha^{1/2}.$ This
means that the critical wavelength reduces directly with the characteristic
thickness of the parallel layers $O(\Ha^{-1/2}).$ The relative phase
velocity for both modes is seen to tend asymptotically to a constant
$c_{c}\approx0.475$ which confirms that this instability is indeed
associated with the sidewall jets and, thus, it is completely determined
by the characteristic thickness and by the velocity of those jets.
Note that the relative phase velocity of two other modes of type II
and IV is also $O(1)$ which implies that these instabilities are
associated with the sidewall jets, too. However, the critical wavenumber
for modes II and IV remains $O(1)$ even for $\Ha\gg1.$ This, in
turn, implies that both of these instability modes are caused by the
velocity variation over the height rather than the thickness of the
jet. Thus, the height rather than thickness of the jet serves as the
characteristic length scale for modes II and IV.

\begin{figure}
\begin{centering}
\includegraphics[bb=150bp 95bp 320bp 255bp,clip,width=0.5\textwidth]{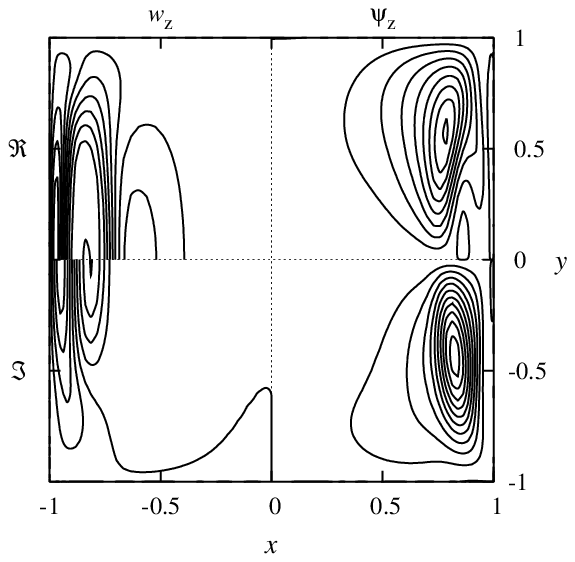}
\par\end{centering}

\caption{\label{fig:egv_Ge1_Ha100v_11} Amplitude distributions of real $(y>0)$
and imaginary $(y<0)$ parts of $\hat{w}$ $(x<0)$ and $\hat{\psi}_{z}$
$(x>0)$ of the critical perturbations over one quadrant of duct cross-section
for instability mode I at $\Ha\approx100$ and $\RE_{c}\approx1170.$}

\end{figure}

Figure \ref{fig:egv_Ge1_Ha100v_11} shows the amplitude distribution
of the most unstable perturbation for type I only because it is almost
identical to that for type III except for the opposite $y$-parity.
In this case, however, the $y$-parity has almost no effect on the
amplitude distributions on each side of the duct because perturbations
are localised in the jets at the side walls and practically do not
interact with each other. As it is seen, for both modes of type I
and III, the component of velocity perturbation along the magnetic
field is an odd function of $y$ whereas the other two velocity components
are even functions. Thus, the transversal circulation in the $(x,y)$-plane
involves a couple of vertically mirror-symmetric vortices at each
sidewall, while the perturbation of the longitudinal velocity $w,$
which is an even function of $y,$ is rather uniform along the magnetic
field in the horizontal mid-part of the duct.

\begin{figure}
\begin{centering}
\includegraphics[width=0.5\textwidth]{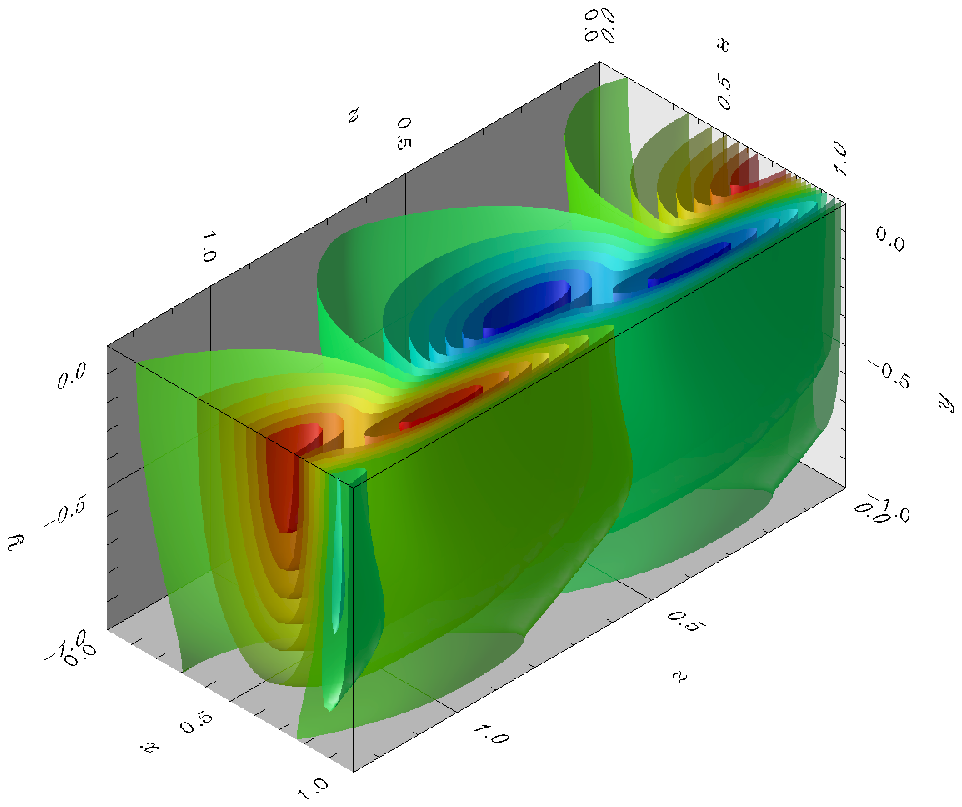}\put(-100,10){(a)}\includegraphics[width=0.5\textwidth]{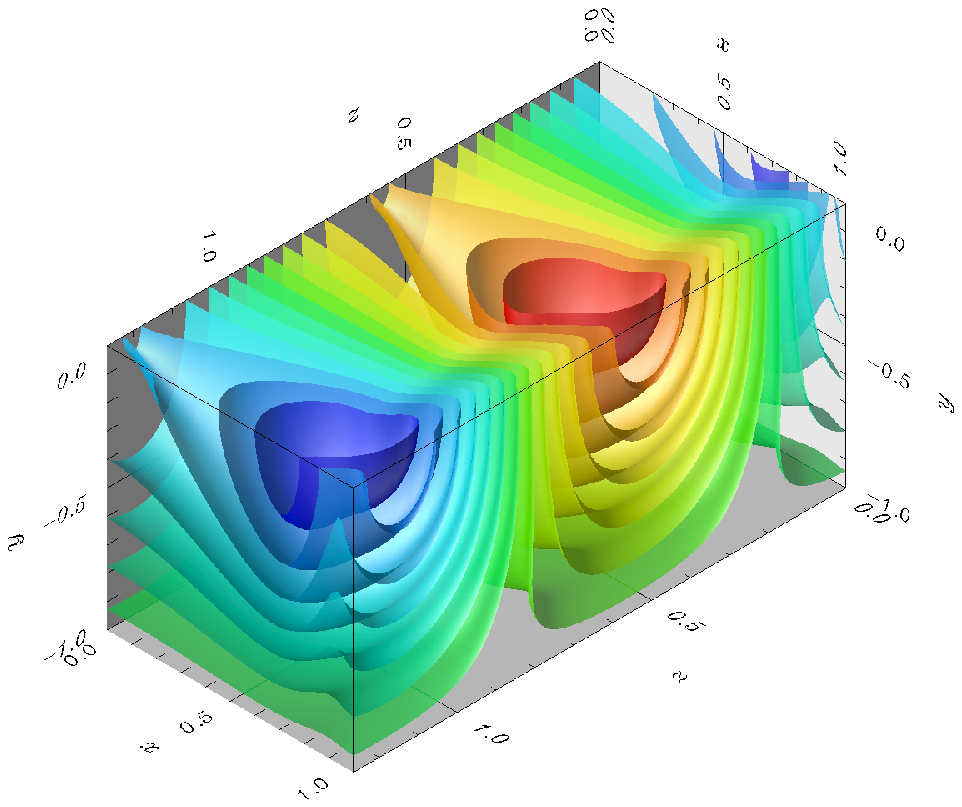}\put(-100,10){(b)}
\par\end{centering}

\caption{\label{fig:sty_Ge1_Ha100v_11} Isosurfaces of $\psi_{y}$ (a) and
electric potential $\phi$ (b) perturbation over a wavelength in one
quadrant of duct cross-section for instability mode I at $\Ha\approx100$
and $\RE\approx1170.$}

\end{figure}

From the energetic point of view, it turns out that $70\%$ and $23\%$
of the kinetic energy are carried by the $z$- and $x$-components
of the velocity perturbation, while only 7\% are carried by the $y$-component.
Thus, $89\%$ of kinetic energy is concentrated in the $y$-component
of vorticity/stream function perturbation, which is associated with
the $z$ and $x$ velocity components. The $x$- and $z$-components
of vorticity/stream function associated with the $y$-component of
velocity contain only $6\%$ and $5\%$ of the energy. Consequently,
in this case, the perturbation of the flow is well represented by
$\psi_{y}$ alone whose isosurfaces, plotted in figure \ref{fig:sty_Ge1_Ha100v_11}(a)
for mode I, show the isolines of solenoidal circulation in the horizontal
plane. The corresponding isosurfaces of the electric potential perturbation
are shown in figure \ref{fig:sty_Ge1_Ha100v_11}(b).

\section{Summary and conclusions}

In this study we have analysed numerically the linear stability of
the flow of a liquid metal in a square duct subject to a transverse
magnetic field. The walls of the duct perpendicular and parallel to
the magnetic field are perfectly conducting and insulating, respectively.
We used a novel 3D vector stream function formulation and Chebyshev
collocation method to solve the eigenvalue problem for small-amplitude
perturbations. Due to the two-fold reflection symmetry of the base
flow with respect to the $x=0$ and $y=0$ planes the perturbations
with four different parity combinations over the duct cross-section
decouple from each other.

The base flow, which without the magnetic field is linearly stable
in a square duct, becomes unstable at  the Hartmann number $\Ha\approx5.7$
as two velocity maxima in the centre of the duct appear. This instability
mode, which is the most dangerous at low Hartmann numbers, involves
the vorticity component in the direction of the magnetic field which
is anti-symmetric and, thus, essentially non-uniform along the field
and symmetric in the spanwise direction across the duct. The velocity
component in the direction of the magnetic field for this mode is
symmetric along the field and anti-symmetric in the spanwise direction,
respectively. This mode becomes the most unstable at $\Ha\approx10$
where its critical Reynolds number based on the maximal velocity attains
a minimum of $\RE_{c}\approx2018.$ The increase of the magnetic field
results in the stabilisation of this mode with $\RE_{c}$ growing
approximately as $\Ha.$ For $\Ha\gtrsim40$ another mode with the
opposite spanwise parity across the duct becomes the most dangerous
and remains such up to $\Ha\approx46.$ These two instability modes
have most of their kinetic energy concentrated in long, streak-like
perturbations of the streamwise velocity which first appear close
to the centre of the duct and then move to the sidewall layers as
the magnetic field increases. The critical wavenumber is $O(1)$ that
corresponds to the critical wavelength considerably exceeding the
width of the duct. It is important to note that the critical phase
velocity remains $O(1)$ even in a relatively strong magnetic field.
This implies that for $\Ha\gg1$ both of these instability modes are
associated with the sidewall jets.

At $\Ha\approx46$ a pair of two additional instability modes appears
with the parity along the magnetic field being opposite to that of
the previous two modes. The critical Reynolds number, which is very
close for both modes, attains a minimum of $\RE_{c}\approx1130$ at
$\Ha\approx70$ and increases as $\RE_{c}\approx91\Ha^{1/2}$ for
$\Ha\gg1.$ The corresponding critical wavelength is $k_{c}\approx0.525\Ha^{1/2}$
while the critical phase velocity approaches $0.475$ of the maximum
jet velocity. This again suggests these two instability modes, similarly
to the first two, to be associated with the side jets. The main difference
between the first and second pairs of disturbances is in their critical
wavenumbers which are $O(1)$ and $O(\Ha^{1/2}),$ respectively. The
latter means that the critical wavelength scales directly with the
side layer thickness $O(\Ha^{-1/2})$, which serves as the characteristic
lengthscale for the last two instability modes. The critical wavenumber
of the first two instability modes being $O(1)$ implies that they
are associated with the velocity variation over the height of the
parallel layer whereas the last two modes are associated with the
velocity variation over the thickness of this layer. From the energetic
point of view, the last two instability modes have most of their kinetic
energy concentrated in the vortical flow component along the magnetic
field which corresponds to the fluid circulation in the planes transverse
to the field.

These last two modes are analogous to the side-layer instability mode
found by \cite{TinWalReePic91} for the flow in the duct with thin
but relatively well-conducting walls. for $\Ha\gg1.$ The critical
Reynolds number based on the average velocity for the latter is $\bar{\RE}_{c}\approx313$
compared to our result $\bar{\RE}_{c}\approx112$ which is rescaled
by the average velocity (\ref{eq:flwr}). On the other hand, our $\bar{\RE}_{c}$
is several times higher than the corresponding result of \cite{Fuj89}
for the two-dimensional approximation. Note that this approximation
incorrectly predicts the base flow in a square duct to remain linearly
unstable in the limit of vanishing magnetic field strength whereas
a significant destabilisation is predicted at the Hartmann numbers
as small as $\Ha\approx10.$ In addition, note that the instability
predicted at $\Ha\approx10$ by our analysis is essentially 3D, as
discussed above, and, thus, it is principally different from the 2D
one found by \cite{Fuj89}. Further comparison with the results of
\cite{TinWalReePic91} shows that our critical wavenumber $\bar{k}_{c}\approx0.525$
scaled by the side-layer thickness $\Ha^{-1/2}$ is close to their
asymptotic value $k_{cr}=0.55.$ At the same time, their phase velocity
$c_{c}=0.0947$ appears to be significantly lower than ours $\bar{c}_{c}=0.423,$
when rescaled with respect to the average velocity. Moreover, the
instantaneous streamlines in the horizontal mid-plane for the critical
perturbation plotted in figure \ref{fig:sty_Ge1_Ha100v_11}(a) show
disconnected sub-vortices at the sidewall whereas those of \cite{TinWalReePic91}
although being similarly deformed are fully connected single vortices.
These differences may be due to the different physical model used
by \cite{TinWalReePic91} as discussed in the Introduction.

In conclusion, note that transiently growing small-amplitude perturbations
may appear below the linear stability threshold due to the so-called
non-normality of the linearised operator (\cite{Tref-etal93}). Transient
growth is sought to account for the bypass transition to turbulence
in the shear flows with a high or none at all linear stability threshold
(\cite{Gro00}). Such a subcritical transition can hardly be relevant
for the Hunt's flow in strong magnetic fields ($\Ha\gtrsim50)$ because
the local critical Reynolds number based on the thickness of side
layers is already very low $\approx100.$ However, it may still be
relevant for weaker magnetic fields, in which the linear stability
threshold is much higher or absent at all, when $\Ha<5.7.$ But even
in the latter case, linear transient growth mechanism might be of
limited importance because, as argued by \cite{Wal95}, \textit{`...the
question of transition is really a question of existence and basin
of attraction of nonlinear self-sustaining solutions that have little
contact with the nonnormal linear problem.}' For a non-magnetic square
duct flow, such nonlinear self-sustaining solutions in the form of
finite amplitude travelling waves have been found recently by \cite{Wed-etal09}
and for the magnetic case by \cite{KinKnaMol09}.

For a flow in the duct with thin conducting walls, the critical Reynolds
numbers observed experimentally by \cite{ReePic89} appear considerably
higher than those predicted by the linear stability theory of \cite{TinWalReePic91}.
This may be owing to the fact that the flow in the experiment was
developing with jets accelerating which would render them more stable,
or that the probes could not reach the thin parallel layer where the
instabilities occur first. However, this may also imply that the side-layer
instability, is supercritical. Then the delay of the transition to
turbulence significantly above the linear stability threshold can
be accounted for by the distinction between the convective and absolute
instabilities. The conventional stability analysis presented in this
paper yields the convective instability threshold at which the flow
becomes able to amplify certain externally imposed perturbations (\cite{Landau87}).
For a small-amplitude supercritical perturbation in the form of travelling
wave to become self-sustained absolute instability is necessary (\cite{LifPit-81}).

\begin{acknowledgements}
The authors are indebted to Leverhulme Trust for financial support
of this work.
\end{acknowledgements}

\end{document}